\documentclass[prd,preprint,superscriptaddress,preprintnumbers,eqsecnum,showpacs,nofootinbib,nobibnotes]{revtex4}
\usepackage{amsfonts,amsmath,amssymb,bm,natbib}
\usepackage{graphicx} 

\newcommand{\be}{\begin{equation}}
\newcommand{\bea}{\begin{eqnarray}}
\newcommand{\ba}{\begin{align}}
\newcommand{\ee}{\end{equation}}
\newcommand{\eea}{\end{eqnarray}}
\newcommand{\ea}{\end{align}}
\def\s#1{{\scriptscriptstyle #1}}

\def\1eq#1{Eq.~(\ref{#1})}

\def\2eqs#1#2{Eqs.~(\ref{#1}) and~(\ref{#2})}
\def\3eqs#1#2#3{Eqs.~(\ref{#1}),~(\ref{#2}) and~(\ref{#3})}
\def\noeq#1{(\ref{#1})}

\def\fig#1{Fig.~\ref{#1}}

\def\BRSTsrc#1{#1^*}
\def\aBRSTsrc#1{#1^\#}
\def\BRSTaBRSTsrc#1{\widehat{#1}}

\def\caBRSTsrc#1#2{#1^{\##2}}
\def\cBRSTaBRSTsrc#1#2{\widehat{#1}^{#2}}
\def\BRSTaBRSTsrcGhost#1{\widehat{#1}}
\def\BRSTaBRSTcovD{\widehat{\cal D}^{ab}_\mu}

\def\wh#1{\widehat{#1}}
\def\G{\Gamma} 

\def\bs{\bar s}

\def\diff{\mathrm{d}}

\begin{document}

\title{AntiBRST symmetry and Background Field Method}
\date{September 3, 2013}

\author{D. Binosi}
\email{binosi@ectstar.eu}
\affiliation{European Centre for Theoretical Studies in Nuclear
  Physics and Related Areas (ECT*) and Fondazione Bruno Kessler, \\ Villa Tambosi, Strada delle
  Tabarelle 286, 
 I-38123 Villazzano (TN)  Italy}
 
\author{A. Quadri}
\email{andrea.quadri@mi.infn.it}
\affiliation{Dip. di Fisica, Universit\`a degli Studi di Milano, via Celoria 16, I-20133 Milano, Italy\\
and INFN, Sezione di Milano, via Celoria 16, I-20133 Milano, Italy}

\begin{abstract}
\noindent

We show that the requirement that a SU(N) Yang-Mills action (gauge fixed in a linear covariant gauge) is invariant under both the Becchi-Rouet-Stora-Tyutin  (BRST) symmetry as well as the corresponding antiBRST symmetry, automatically implies that the theory is quantized in the (linear covariant) background field method (BFM) gauge. Thus, the BFM and its associated background Ward identity naturally emerge from antiBRST invariance of the theory and need not be introduced as an {\it ad hoc} gauge fixing procedure. Treating ghosts and antighosts on an equal footing, as required by a BRST-antiBRST invariant formulation of the theory, gives also rise to a local antighost equation that together with the local ghost equation completely resolve the algebraic structure of the ghost sector for any value of the gauge fixing parameter. We finally prove that the background fields are stationary points of the background effective action obtained when the quantum fields are integrated out. 

\end{abstract}
\pacs{11.15.Tk, 12.38.Aw}

\maketitle

\section{Introduction}

Quantization of gauge theories in the presence of background field configurations~\cite{DeWitt:1967ub,Honerkamp:1972fd,Kallosh:1974yh,KlubergStern:1974xv,Arefeva:1974jv,Hooft:1975vy,Weinberg:1980wa,Shore:1981mj,Abbott:1983zw,Hart:1984jy,Abbott:1980hw,Abbott:1981ke} is known to be a very useful tool, for it allows to preserve gauge invariance with respect to external background sources after a gauge-fixing choice has been made for the quantum gauge modes of the theory.

This has led to a number of applications, both at the perturbative level -- ranging from calculations in Yang-Mills theories~\cite{Abbott:1980hw,Ichinose:1981uw} via the quantization of the Standard Model~\cite{Denner:1994xt} to gravity and supergravity calculations~\cite{Gates:1983nr} -- as well as at the non-perturbative level -- where the method has been instrumental in devising a gauge invariant truncation scheme for the Schwinger-Dyson equations of Yang-Mills theories~\cite{Aguilar:2006gr,Binosi:2007pi,Binosi:2008qk,Binosi:2009qm}.

A common algebraic framework has emerged over the years in order to tame the dependence of the vertex functional $\G$ on the background fields. In~\cite{Grassi:1995wr} it was first proposed to introduce a BRST partner $\Omega$ for the background field $\widehat A$. The corresponding extended Slavnov-Taylor (ST) identity guarantees that the physics (described by the cohomology of the linearized ST operator) is not affected by the introduction of the background. This approach allows to acquire an algebraic control over the renormalization of the theory under scrutiny~\cite{Grassi:1997mc,Grassi:1999nb} and to prove
the so-called Background Equivalence Theorem~\cite{Becchi:1999ir,Ferrari:2000yp}.   

Eventually it has been recognized in a series of papers~\cite{Binosi:2011ar,Binosi:2012pd,Binosi:2012st} that the full dependence on the background field, fixed by the extended ST identity, is induced through a canonical transformation 
with respect to (w.r.t.) the Batalin-Vilkovisky (BV) bracket of the theory. Such a canonical transformation is generated by the functional ${\delta \G}/{\delta \Omega}$. Since the latter is in general background-dependent, one cannot obtain the finite form of the canonical transformation by simple exponentiation; rather, one needs to resort to the Lie series of an appropriate functional differential operator. The derivation proceeds in close analogy with the case of parameter-dependent canonical transformations in classical mechanics~\cite{Binosi:2012st}.

In this paper this simple geometrical interpretation of the background field method will be pushed one step further, as a very deep connection between the BFM and the so-called antiBRST symmetry will be unveiled.

Indeed, since the advent of BRST quantization, it has been known that a further symmetry exists, induced by an antiBRST transformation~\cite{Curci:1976bt,Ojima:1980da,Baulieu:1981sb} in which the antighost field takes the place of the ghost in the variation of the gauge and matter fields of the theory. Though this symmetry turned out to be a useful tool for constraining possible terms in the action and in simplifying relations between Green's functions, it has however been far from clear if there is any case in which it is indispensable, and thus its meaning has remained so far somewhat mysterious.

The antiBRST symmetry generates an antiST identity, that can be shown to hold together with the ST identity induced by the BRST transformations. Moreover, the requirement of simultaneous BRST and antiBRST invariance can be fulfilled, so that both identities hold true for the vertex functional $\G$, provided that a suitable set of operators for the BRST-antiBRST variation of the fields is introduced through the coupling to appropriate external classical sources. As we will see {\it the latter sources coincide precisely with the background fields introduced in the BFM}.
Thus, for example, the BRST partner $\Omega$ of the background gauge field $\widehat A$ is seen to be the antifield of the antiBRST transformation $\bar s A$ for the gauge field $A$, while $\wh{A}$ is the source coupled to the
BRST-antiBRST variation $s \bar s A$ of $A$.

Similar identifications hold true for all the other scalar and fermionic matter fields of the theory.
In particular, we will show that  for spontaneously broken gauge theories the procedure automatically yields
the correct background 't Hooft gauge-fixing.

In addition, in a BRST-antiBRST invariant theory both a local ghost and a local antighost equation exist. While the former equation has been known since a long time, the latter has been derived up to now only in the background Landau gauge~\cite{Grassi:2004yq}. However, once the sources required to establish the ST and the antiST identities are introduced, it can be readily seen that there is nothing special in the Landau gauge choice and one can indeed construct a local antighost equation valid for a general $R_\xi$-gauge. The usual background Ward identity arises then both as a consequence of the validity of the ST identity and the local antighost equation, as well as
of the antiST identity and the local ghost equation.

As our analysis reveals that the background field method is naturally encoded in every theory which is both BRST and antiBRST invariant, one might wonder whether the ST and the antiST identities impose some physical condition on the backgrounds. For that purpose, it is convenient to construct an effective action $\widetilde \G$ for the background fields, where the quantum modes have been fully integrated out, {\it i.e.}, one keeps connected graphs with external backgrounds only (and therefore $\widetilde \G$ is one-particle reducible w.r.t. the quantum fields). This approach is motivated by several applications of the BFM, {\it e.g.}, in the effective field theory of the Color Glass Condensate~\cite{Hatta:2005rn,Iancu:2000hn,Ferreiro:2001qy}.

As we will see, the ST and the antiST identities imply then that the background field configurations are a stationary point for $\widetilde \G$. This leads to some interesting combinatorial relations of a novel type between one-particle irreducible (1-PI) graphs involving external background sources, whose origin can be ultimately traced back to the canonical transformation of~\cite{Binosi:2012st}, which dictates the dependence of $\G$ on the backgrounds. 

The paper is organized as follows.
In Section~\ref{sec:gauge-fixing} we discuss the conventional
and the background ($R_\xi$) gauge-fixing for a pure SU(N) Yang-Mills theory
and derive the associated extended ST identity.
In Section~\ref{sec:antiBRST} the antiBRST symmetry is introduced together
with the sources required to define the composite operators of the
BRST-antiBRST algebra. The equivalence between the BRST-antiBRST
invariance and the BFM is then established. Next, 
Section~\ref{sec:ag} is dedicated to the derivation of the local antighost equation
in a generic $R_\xi$ gauge. In addition, we show how the background Ward identity emerges from both the antiST identity combined with the local ghost equation
or the ST identity combined with the local antighost equation.
The local ghost and antighost equations are then exploited to fully constrain the ghost two-point sector in any gauge. In Section~\ref{sec:BEA} we finally construct the background effective action by integrating out all quantum fields, and show that the background fields are stationary points of this action. Our conclusions are presented in Section~\ref{sec:concl}, with the following Appendix generalizing (some of) the main equations for the case in which scalars and fermion fields are present.  

\section{Conventional and BFM gauge-fixings\label{sec:gauge-fixing}}

The action of a SU(N) Yang-Mills theory reads
\be
S=S_\s{\mathrm{YM}}+S_\s{\mathrm{GF}}+S_\s{\mathrm{FPG}}.
\label{cl-action}
\ee
$S_\s{\mathrm{YM}}$ represents the Yang-Mills (gauge invariant) action, 
which is written in terms of the SU(N) field strength
\be
S_\s{\mathrm{YM}}
=-\frac1{4g^2}\int\! \diff^4x\,F_{\mu\nu}^a F^{\mu\nu}_a;\qquad
F_{\mu\nu}^a=\partial_\mu A^a_\nu-\partial_\nu A^a_\mu+f^{abc}A^b_\mu A^c_\nu.
\ee
$S_\s{\mathrm{GF}}$ and $S_\s{\mathrm{FPG}}$ represent respectively the (covariant) gauge-fixing functional and its associated Faddeev-Popov ghost term. The most general way of writing these terms is through the expressions
\be
S_\s{\mathrm{GF}}=\int\! \diff^4x\,\left[-\frac\xi2(b^a)^2+b^a{\cal F}^a\right];\qquad
S_\s{\mathrm{FPG}}=-\int\! \diff^4x\,\bar c^a s {\cal F}^a.
\ee
In the formulas above ${\cal F}$ represents the gauge-fixing function, which, for the class of $R_\xi$ gauges considered throughout this paper, reads
\be
{\cal F}^a=\partial^\mu A^a_\mu.
\label{Rxi-gf}
\ee
In addition, $b$ are auxiliary, non-dynamical fields (the so called Nakanishi-Lautrup multipliers) that can be eliminated through their equations of motion, 
as a consequence of the validity of the $b$-equation
\be
\frac{\delta \Gamma}{\delta b^a} = - \xi b^a + {\cal F}^a,
\ee
for the full quantum effective action $\G$.
$c$ (respectively, $\bar c$) are the ghost~(respectively, antighost) fields, while, finally, $s$ is the nilpotent BRST operator, which constitutes a symmetry of the gauge-fixed action~\1eq{cl-action}, with the BRST transformations of the various fields given by
\be
sA^a_\mu={\cal D}^{ab}_\mu c^b;\qquad
sc^a=-\frac12f^{abc}c^bc^c;\qquad
s\bar c^a=b^a;\qquad
sb^a=0,
\label{BRST}
\ee
and the covariant derivative ${\cal D}$ is defined according to
\be
{\cal D}^{ab}_\mu=\partial_\mu\delta^{ab}+f^{acb}A^c_\mu.
\ee
We thus see that the sum of the gauge-fixing and Faddeev-Popov terms can be written as a total BRST variation:
\be
S_\s{\mathrm{GF}}+S_\s{\mathrm{FPG}}=\int\! \diff^4x\,s\left[\bar c^a{\cal F}^a-\frac\xi2\bar c^ab^a\right].
\label{eq.GF.plus.FPG}
\ee
This is of course expected, for it is well known that the physical observables of a theory admit a mathematical characterization in terms of the local cohmology of the BRST operator~\cite{Barnich:2000zw,Barnich:1994mt,Barnich:1994db}, and the latter is not affected by total BRST variations.
 
The BRST symmetry of the Yang-Mills action can be most conveniently exposed thorugh the so-called Batalin-Vilkoviski (BV) method, {\it i.e.}, introducing a set of antifields $\Phi^*$ and coupling them to the BRST variation of the corresponding fields through the term~\cite{Gomis:1994he}
\be
S_\s{\mathrm{BV}}=\int\! \diff^4x\sum\Phi^*s\,\Phi.
\ee
Then the (tree-level) vertex functional is given by the sum
\be
\Gamma^{(0)}=S_\s{\mathrm{YM}}+S_\s{\mathrm{GF}}+S_\s{\mathrm{FPG}}+S_\s{\mathrm{BV}},
\ee
and the BRST symmetry of the action is encoded by the ST identity
\be
\int\! \diff^4x\left[\frac{\delta\Gamma}{\delta A^{*a}_\mu}\frac{\delta\Gamma}{\delta A^{\mu}_a}+\frac{\delta\Gamma}{\delta c^{*}_a}\frac{\delta\Gamma}{\delta c^a}+b^a\frac{\delta\Gamma}{\delta \bar c^{a}}\right]=0,
\label{conv-ST-identity}
\ee
where now $\Gamma$ is the full quantum effective action.

Turning to the case of BFM type of gauges, traditionally  one starts by splitting the gauge field into a background part ($\widehat{A}$) and a quantum part ($Q$) according to
\be
A^a_\mu=\widehat{A}^a_\mu+Q^a_\mu.
\label{cl-split}
\ee
Next, one retains the background gauge invariance of the gauge-fixed action by choosing a gauge-fixing function that transforms in the adjoint representation of SU(N) through the general replacements
\be
\partial_\mu\delta^{ab}\to\widehat{\cal D}^{ab}_\mu\equiv\partial_\mu\delta^{ab}+f^{acb}\widehat{A}^c_\mu;\qquad
A^a_\mu\to Q^a_\mu,
\ee
that is one has the background $R_\xi$ gauge
\be
\widehat{\cal F}^a=\widehat{\cal D}^{ab}_\mu Q^\mu_b.
\label{BFM-gf}
\ee
Finally, in addition to the anti-fields $\Phi^*$, the quantization of the theory in the BFM requires~ the introduction of an additional (vector) anticommuting source $\Omega$,  implementing the equation of motion of the background field at the quantum level, with~\cite{Grassi:1995wr}
\be
s\widehat{A}^a_\mu=\Omega^a_\mu;\qquad s\,\Omega^a_\mu=0.
\label{Omegadoublet}
\ee
The BRST transformation of the quantum field $Q$ is given by
\be
sQ^a_\mu={\cal D}^{ab}_\mu c^b-\Omega^a_\mu.
\ee
\1eq{Omegadoublet} ensures that $\widehat{A}$ and $\Omega$ are paired in a so-called BRST-doublet~\cite{Barnich:2000zw,Quadri:2002nh} (as already happens for $\bar c$ and $b$), thus preventing the background field from modifying the physical observables of the theory.

It then follows that the conventional ST identity~\1eq{conv-ST-identity} gets modified into the extended ST identity
\be
\int\! \diff^4x\left[\frac{\delta\Gamma}{\delta A^{*a}_\mu}\frac{\delta\Gamma}{\delta Q^{\mu}_a}+\frac{\delta\Gamma}{\delta c^{*a}}\frac{\delta\Gamma}{\delta c^a}+b^a\frac{\delta\Gamma}{\delta \bar c^{a}}+\Omega^\mu_a\left(\frac{\delta\Gamma}{\delta \widehat{A}_{\mu}^a}-\frac{\delta\Gamma}{\delta Q_{\mu}^a}
\right)\right]=0.
\ee
By ``undoing'' the shift of the gauge field~\noeq{cl-split} the ST identity above may be cast in the somewhat more compact form
\be
\int\! \diff^4x\left[\frac{\delta\Gamma}{\delta A^{*a}_\mu}\frac{\delta\Gamma}{\delta A^{\mu}_a}+\frac{\delta\Gamma}{\delta c^{*}_a}\frac{\delta\Gamma}{\delta c^a}+b^a\frac{\delta\Gamma}{\delta \bar c^{a}}+\Omega^\mu_a\frac{\delta\Gamma}{\delta \widehat{A}_{\mu}^a}\right]=0.
\ee
In particular, this formulation of the BFM in terms of the field variables $A$ and $\widehat{A}$ turns out to be the most suitable for the ensuing analysis.

\section{Anti-BRST symmetry and the BFM\label{sec:antiBRST}}

In the BRST transformations the role of the ghost field $c$ is very prominent, as it replaces the gauge transformation parameter  of the conventional gauge transformations and its behavior can be understood in an intrinsic manner in terms of the cohomology of the Lie algebra~(see, {\it e.g.},~\cite{Nakanishi:1977ae,Barnich:2000zw}). On the other hand, the antighost $\bar c$ and its doublet partner $b$ play the  role of Lagrange multipliers introduced to enforce the gauge-fixing condition ${\cal F}=0$ and its BRST transform $s{\cal F}=0$. In addition, $\bar c$ obeys an equation of motion which is different from that of $c$ as the former is not the hermitian conjugate of the latter field.

Though all these seemed  to rule out the possibility that $\bar c$ and $c$  can be interchanged, a nilpotent `antiBRST' transformation symmetry in which this is exactly what happens was introduced long ago~\cite{Curci:1976bt,Ojima:1980da,Baulieu:1981sb}.  
 Indeed, the antiBRST transformations can be obtained from the BRST ones of~\1eq{BRST} by exchanging the role of the ghost and antighost fields; that is one has
\be
\bs A^a_\mu={\cal D}^{ab}_\mu \bar c^b;\qquad
\bs \bar c^a=-\frac12f^{abc}\bar c^b\bar c^c;\qquad
\bs c^a=\bar b^a;\qquad
\bs \bar b^a=0.
\label{antiBRST}
\ee
In particular, the antiBRST transformation of the gauge field is
obtained from the gauge variation of $A$ by replacing the gauge parameter
by the antighost field $\bar c$. 

In order to close the algebra the transformations above need to be supplemented with the additional transformations
\be
s\bar b^a=f^{abc}\bar b^b c^c;\qquad
\bs b^a=f^{abc} b^b\bar c^c.
\label{additional}
\ee

On the other hand, as both $s$ and $\bs$ are nilpotent, the additional (natural) requirement that their sum is also nilpotent (or that $\{s,\bs\}=0$), results in the constraint~\cite{Curci:1976bt}
\be
\bar b^a=-b^a-f^{abc}c^b\bar c^c,
\label{constr}
\ee
which, upon use of the Jacobi identity, is readily seen to be consistent with 
\1eq{additional}. 

Finally, the nontrivial BRST-antiBRST transformations of the fields read
\be
s\bs A^a_\mu={\cal D}^{ab}_\mu b^b+f^{abc}\left({\cal D}_\mu^{bd}c^d\right)\bar c^c;\qquad
s\bs c^a=s\bar b^a;\qquad
s\bs \bar c^a=-\bs b^a.
\ee

At this point it is straightforward to realize that to render our theory~\noeq{cl-action} simultaneously BRST and antiBRST invariant, requires,  {\it before gauge-fixing}, the introduction of 8 sources: the usual antifields $A^*$ and $c^*$, the antiBRST sources $\aBRSTsrc{A}$, $\aBRSTsrc{c}$, $\aBRSTsrc{\bar c\hspace{0.05cm}}$ and $\aBRSTsrc{b}$, and, finally, the BRST-antiBRST sources $\BRSTaBRSTsrc{A}$ and $\BRSTaBRSTsrcGhost{c}$. Notice that we do not add any source associated to $s\bar b$, for, due to the constraint~\noeq{constr}, the BRST transformation of this field can be completely recovered from the corresponding transformations of $b$, $c$, and $\bar c$.
One has then that the BRST-antiBRST invariant action reads
\be
S_\s{\mathrm I}=S_\s{\mathrm{YM}}+\sum\!\int\mathrm{d}^4x\,\left(\Phi^*s\,\Phi+\aBRSTsrc{\Phi}\bar s\,\Phi+\BRSTaBRSTsrc{\Phi}s\bar s\,\Phi\right),
\label{inv-action}
\ee
where the sum extends over all the nonzero sources, and  (with the exception of $\aBRSTsrc{b}$)
\begin{align}
s\,\Phi^*&=\bar s\,\Phi^*=0;& 
s\,\BRSTaBRSTsrc{\Phi}&=\aBRSTsrc{\Phi};\nonumber \\
s\,\aBRSTsrc{\Phi}&=\bar s\,\aBRSTsrc{\Phi}=0;&
\bar s\,\BRSTaBRSTsrc{\Phi}&=-\Phi^*.
\label{brst.abrst.srcs}
\end{align}
For the source $\aBRSTsrc{b}$ one has instead 
\be
s\,\aBRSTsrc{b}_a=\aBRSTsrc{\bar c}_a; \qquad \bs\,\aBRSTsrc{b}_a=0.
\ee
Finally, the  ghost charge assignments are
\begin{align}
\mathrm{gh}(\Phi^*)&=-\mathrm{gh}(\Phi)-1;&
\mathrm{gh}(\aBRSTsrc{\Phi})&=-\mathrm{gh}(\Phi)+1;&
\mathrm{gh}(\BRSTaBRSTsrc{\Phi})&=-\mathrm{gh}(\Phi),&
\end{align}
where we have set $\mathrm{gh}(c,\bar c)=(1,-1)$. Notice that the usual BV action~\cite{Gomis:1994he} is recovered by setting the $\aBRSTsrc{\Phi}$ and $\BRSTaBRSTsrc{\Phi}$ sources to zero.

We are now ready to establish the central result of this paper. Consider, in fact, the BFM covariant gauge-fixing~\noeq{BFM-gf} with its associated Faddeev-Popov ghost action; a straightforward calculation yields
\bea
s\left[\bar c^a\widehat{\cal F}^a-\frac\xi2\bar c^a b^a\right]&=&s\left[\bar c^a{\cal F}^a-\frac\xi2\bar c^a b^a\right]+\widehat{A}^\mu_a({\cal D}^{ab}_\mu b^b+f^{abc}\bar c^b{\cal D}_\mu^{cd}c^d)+\Omega^\mu_a({\cal D}^{ab}_\mu\bar c^b)\nonumber\\
&=&s\left[\bar c^a{\cal F}^a-\frac\xi2\bar c^a b^a\right]+\widehat{A}^\mu_a (s\bar s A^a_\mu)+\Omega^\mu_a(\bar s A^a_\mu),
\label{main}
\eea
where ${\cal F}^a$ is now the covariant gauge-fixing~\noeq{Rxi-gf}. As a result of the anticommutation relation $\{s,\bar s\}=0$ and the identity
\be
s(\bar c^a\partial^\mu A^a_\mu)=-\bar s(c^a\partial^\mu A^a_\mu),
\ee
we observe that also the first term in the right-hand side (r.h.s.) of~\1eq{main} is  both BRST and antiBRST invariant (the term $b^2$ is obviously invariant under these transformations).

Then we see that by adding to the  BRST-antiBRST invariant action~\noeq{inv-action} the $R_\xi$ gauge-fixing and Faddev-Popov term~(\ref{eq.GF.plus.FPG}) we automatically obtain a theory formulated in the background $R_\xi$ gauge, provided that the following identification is made:
\be
\Omega^a_\mu\equiv \caBRSTsrc{A}{a}_\mu \, .
\label{ids}
\ee
From \1eq{main} one also sees that the background gauge field $\wh{A}$ is the source
of the BRST-antiBRST variation $s \bar s A$ of the gauge field $A$. 
Notice however that $\widehat{c}$ cannot be interpreted as a background 
for the ghost $c$, since it has ghost number $-1$; 
it is also clear that it is not a background for the antighost field, as a shift of the latter field would lead to totally different couplings w.r.t. the ones that are generated for the source $\widehat{c}$.

Thus one arrives at the somewhat surprising conclusion that {\it requiring the invariance of a  {\rm SU(N)} Yang-Mills  action gauge-fixed in an $R_\xi$ gauge under both BRST as well as antiBRST symmetry is equivalent to quantizing the theory in the ($R_\xi$) BFM}:
\be
\Gamma^{(0)} =S_\s{\mathrm{I}}+S_\s{\mathrm{GF}}+S_\s{\mathrm{FPG}}  = S_\s{\mathrm{YM}}+\widehat{S}_\s{\mathrm{GF}}+\widehat{S}_\s{\mathrm{FPG}} + S_\s{\mathrm{BV}}
 + \int \diff^4x \, \left ( \aBRSTsrc{c}_a \bar s c_a + \BRSTaBRSTsrcGhost{c}_a 
 s \bar s c_a \right ) ,
\label{eq.bfm-der}
\ee
where the background gauge-fixing functional
and the background Faddeev-Popov terms are
\be
\widehat{S}_\s{\mathrm{GF}}=\int\! \diff^4x\,\left[-\frac\xi2(b^a)^2+b^a\widehat{\cal F}^a\right];\qquad
\widehat{S}_\s{\mathrm{FPG}}=-\int\! \diff^4x\,\bar c^a s \widehat{\cal F}^a.
\ee
The standard BFM tree-level vertex functional is recovered
by setting $\aBRSTsrc{c} = \wh{c}=0$ in the r.h.s. of~\1eq{eq.bfm-der}.
In this sense the BFM is not fundamental, as it is naturally emerging from the requirement of antiBRST invariance.

It is interesting to study the case in which  (complex) scalars and/or fermions are added to the theory. Let's start from the former fields, where one has
\begin{align}
s\,\phi&=ic^a t^a\phi;& 
\bar s\,\phi&=i\bar c^at^a\phi;&
s\phi^\dagger&=-ic^a\phi^\dagger t^a;&
\bar s\phi^\dagger&=-i\bar c^a\phi^\dagger t^a,&
\label{sc-trans}
\end{align}
with $t^a$  the generators of the SU(N) representation chosen for $\phi$. 
The corresponding BRST-antiBRST transformation reads
\be
s\bar s\,\phi=ib^at^a\phi+\bar c^a c^b t^a t^b\phi;\qquad
s\bar s\,\phi^\dagger=-ib^a\phi^\dagger t^a+\bar c^a c^b\phi^\dagger t^b t^a,
\label{sc-2trans}
\ee
from which it is immediate to infer that $\{s,\bar s\}\,\phi=0$.

The extra sources one needs to add to render the action BRST and antiBRST invariant in the presence of the scalar field $\phi$ are then\footnote{We assume that a suitable (gauge invariant) action term $S_\phi$ (and $S_\psi$ when adding fermions) is added to the classical action~\noeq{cl-action}; its concrete form is however irrelevant for the following analysis.}
\be
\int\!\diff^4x\,\left[{\BRSTsrc{\phi}}^\dagger s \phi + s \phi^\dagger \BRSTsrc{\phi}+\caBRSTsrc{\phi}{\dagger}\bar s\,\phi+(\bar s\,\phi^\dagger)\aBRSTsrc{\phi}+\cBRSTaBRSTsrc{\phi}{\dagger}s\bar s\,\phi+(s\bar s\,\phi^\dagger)\BRSTaBRSTsrc{\phi}\right] .
\ee

Again by identifying $\wh{\phi}$ and $\wh{\phi}^\dagger$ with
the background for the scalars $\phi$ and $\phi^\dagger$ respectively,
as well as
$\aBRSTsrc{\phi},\caBRSTsrc{\phi}{\dagger}$
with their corresponding BRST doublet partners ({\it i.e.},
$s \phi = \aBRSTsrc{\phi}, ~ s \phi^\dagger = \caBRSTsrc{\phi}{\dagger}$,
as prescribed by~\1eq{brst.abrst.srcs}), 
%
%
%
one recovers the background~'t~Hooft gauge after the background field $\widehat{\phi}$ has acquired an expectation value~$v$.

For fermions $\psi$ and $\bar\psi$ the analysis proceeds in the same way as in the scalar case, since \2eqs{sc-trans}{sc-2trans} still hold once $t^a$ is identified with the generator of the representation of the fermionic matter field and $\phi$ replaced by $\psi$, {\it i.e.},
\begin{align}
s\psi&=ic^at^a\psi;&
\bar s\psi&=i\bar c^at^a\psi;&
s\bar\psi&=-ic^a\bar\psi t^a;&
\bar s\bar\psi&=-i\bar c^a\bar\psi t^a.
\end{align}
Notice that the requirement of antiBRST invariance generates unavoidably the sources $\BRSTaBRSTsrc{\psi}$ and $\BRSTaBRSTsrc{\bar\psi}$ which correspond to background fields for the fermions\footnote{Fermionic backgrounds have been considered, {\it e.g.}, in~\cite{Jack:1984vj}; their physical relevance is however unclear to us at the moment.}, as the action will be rendered BRST-antiBRST invariant through the addition of the term
\be
\int\!\diff^4x\,\left[\BRSTsrc{\bar \psi} s\psi - (s \bar \psi) \BRSTsrc{\psi} +\aBRSTsrc{\bar\psi}\bs\,\psi-(\bs\,\bar\psi)\aBRSTsrc{\psi}
-\BRSTaBRSTsrc{\bar\psi}s\bs\,\psi+(s\bs\,\bar\psi)\BRSTaBRSTsrc{\psi}\right], 
\ee 
where
\be
s\bar s\,\psi=ib^at^a\psi+\bar c^a c^b t^a t^b\psi;\qquad
s\bar s\,\bar\psi=-ib^a\bar\psi t^a+\bar c^a c^b\bar\psi t^b t^a.
\label{fer-2trans}
\ee

\section{Local antighost equation}\label{sec:ag}

The presence of the antiBRST symmetry leads, as we will explicitly show below,  to the existence of a local antighost equation. It should be noticed that for a SU($N$) Yang-Mills theory with a conventional $R_\xi$ gauge-fixing only an {\it integrated} antighost equation can be derived, while in the BFM case the existence of a local version of this equation was established in the background Landau gauge in~\cite{Grassi:2004yq} and believed to be valid only for that specific gauge-fixing choice.

On the other hand, the correspondence just found between BRST and antiBRST invariance and the BFM shows that there should not be anything special neither when formulating the theory in the BFM nor when choosing $\xi=0$. Indeed as the existence of the antiBRST symmetry puts the ghost and antighost fields on the same footing, and given that a local ghost equation (sometimes also referred to as Faddeev-Popov equation) is known to hold, we would expect a local antighost equation to hold as well.  

To show that this is indeed the case, let us  start by setting to zero the scalar and fermionic matter sector (the complete case will be discussed in Appendix~\ref{app:A}); then  the tree-level action~\noeq{inv-action} can be cast in the form
\be
\Gamma^{(0)}=S_\s{\mathrm{YM}}+ s X=S_\s{\mathrm{YM}}+
 \bar s\, Y,
\ee
where
\bea
X&=& \int \diff^4 x\, \left [ \sum\left((-1)^{\mathrm{gh}(\Phi^*)}\Phi^*\Phi+\BRSTaBRSTsrc{\Phi}\bar s \Phi\right)+\bar c^a{\cal F}^a-\frac\xi2\bar c^ab^a \right ],
\nonumber \\
Y&=& \int \diff^4x \, \left [ \sum\left((-1)^{\mathrm{gh}(\aBRSTsrc{\Phi})}\aBRSTsrc{\Phi}\Phi+\BRSTaBRSTsrc{\Phi} s \Phi\right)-c^a{\cal F}^a+\frac\xi2 c^ab^a \right ],
\label{XY}
\eea
and the sum is intended, as familiar by now,  over all nonzero sources. 


To derive the local antighost equation the fastest route turns out to be to calculate the anticommutator between the derivative w.r.t. the ghost field and the antiBRST operator.
Since
\be
\bar s
=\sum \int \diff^4 x\, ~ \bar s\,\varphi(x)\frac{\delta}{\delta\varphi(x)};\qquad\varphi={\Phi,\Phi^*,\aBRSTsrc{\Phi},\BRSTaBRSTsrc{\Phi}},
\ee
one finds that for any functional $F=F[\varphi]$ with zero ghost charge
\be
\left\{\frac{\delta}{\delta c^a},\bar s\right\}F=\sum
\int \diff^4x \,
\left[
\frac{\delta}{\delta c^a}\bar s\,\varphi(x)\right]\frac{\delta F}{\delta\varphi(x)}.
\ee 
and therefore
\be 
\frac{\delta\Gamma}{\delta c^a}=
\frac{\delta}{\delta c^a}(\bar s\, Y)=\sum \int \diff^4 x \, \left[\frac{\delta}{\delta c^a}\bar s\,\varphi( x)\right]\frac{\delta Y}{\delta\varphi(x)}-
\bar s\,\frac{\delta Y}{\delta c^a}.
\ee

Then through a lengthy but relatively straightforward calculation, we arrive at the local antighost equation
\bea
\bar {\cal G}_a \Gamma &\equiv& \frac{\delta\Gamma}{\delta c^a} 
+ f^{abc}\frac{\delta\Gamma}{\delta b^b}\bar c^c+\xi\frac{\delta\Gamma}{\delta \aBRSTsrc{b}_a}-\BRSTaBRSTcovD\frac{\delta\Gamma}{\delta \caBRSTsrc{A}{b}_\mu}
-f^{abc}\BRSTaBRSTsrcGhost{c}_b\frac{\delta\Gamma}{\delta \aBRSTsrc{c}_c}-f^{abc}\aBRSTsrc{b}_b\frac{\delta\Gamma}{\delta \aBRSTsrc{\bar c\hspace{0.05cm}}_c} \nonumber \\
&=& {\cal D}^{ab}_\mu A^{*\mu}_b +f^{abc}c^*_bc^c,
\label{AGE-xi}
\eea
%
where $\BRSTaBRSTcovD=\partial_\mu\delta^{ab}+f^{acb}\cBRSTaBRSTsrc{A}{c}_\mu$. Notice that all the (possibly present) trilinear terms in the ghost and antighost fields have cancelled out.

In the case of the local ghost equation one computes the anticommutator of the derivative w.r.t. the antighost field and the BRST operator $s$:
\be
\frac{\delta\Gamma}{\delta \bar c^a}=\frac{\delta}{\delta \bar c^a}(s X)=\sum
\int \diff^4x \, \left[\frac{\delta}{\delta \bar c^a}s\,\varphi(x)\right]\frac{\delta X}{\delta\varphi(x)}-s\,\frac{\delta X}{\delta \bar c^a}. 
\ee
One then has
\bea
{\cal G}_a \Gamma &\equiv& \frac{\delta\Gamma}{\delta \bar c^a} 
+\BRSTaBRSTcovD\frac{\delta\Gamma}{\delta A^{*b}_\mu}+f^{abc}\BRSTaBRSTsrcGhost{c}_b\frac{\delta\Gamma}{\delta c^*_c} \nonumber \\
&=&{\cal D}^{ab}_\mu \caBRSTsrc{A}{\mu}_b
+f^{abc} \aBRSTsrc{c}_b c^c+f^{abc}\aBRSTsrc{\bar c\hspace{0.05cm}}_b\bar c^c-f^{abc}\aBRSTsrc{b}_bb^c.
\label{GE-xi}
\eea

Finally, the $b$ equation assumes the form
\bea
\frac{\delta\Gamma}{\delta b^a}=\BRSTaBRSTcovD(A^\mu_b-\cBRSTaBRSTsrc{A}{\mu}_b)-\xi b^a-f^{abc}\aBRSTsrc{b}_b\bar c^c-\aBRSTsrc{c}_a-f^{abc}\BRSTaBRSTsrcGhost{c}_bc^c,
\label{bE}
\eea
while the ST and antiST identities read respectively\footnote{From here we see that an alternative way of deriving the local antighost equation is to take the derivative w.r.t. $b$ of the antiST identity~\noeq{antiST-id} and next use the $b$ equation~\noeq{bE} to replace the various terms  involving the functional derivative w.r.t.~$b$.}
\bea
&&\int\!\diff^4x\left[\frac{\delta\Gamma}{\delta A^{*a}_\mu}\frac{\delta\Gamma}{\delta A^{\mu}_a}+\frac{\delta\Gamma}{\delta c^*_a}\frac{\delta\Gamma}{\delta c^a}+\caBRSTsrc{A}{a}_\mu\frac{\delta\Gamma}{\delta\cBRSTaBRSTsrc{A}{a}_\mu}+\aBRSTsrc{c}_a\frac{\delta\Gamma}{\delta\BRSTaBRSTsrcGhost{c}_a}+\aBRSTsrc{\bar c}_a\frac{\delta\Gamma}{\delta\aBRSTsrc{b}_a}+
b^a\frac{\delta\Gamma}{\delta\bar c^a}
\right]=0,
\label{ST-id}\\
&&\int\!\diff^4x\left[\frac{\delta\Gamma}{\delta \caBRSTsrc{A}{a}_\mu}\frac{\delta\Gamma}{\delta A^{\mu}_a}+\frac{\delta\Gamma}{\delta \aBRSTsrc{c}_a}\frac{\delta\Gamma}{\delta c^a}+\frac{\delta\Gamma}{\delta \aBRSTsrc{\bar c\hspace{0.05cm}}_a}\frac{\delta\Gamma}{\delta \bar c^a}+\frac{\delta\Gamma}{\delta \aBRSTsrc{b}_a}\frac{\delta\Gamma}{\delta b^a}-A^{*a}_\mu\frac{\delta\Gamma}{\delta\cBRSTaBRSTsrc{A}{a}_\mu}-c^{*}_a\frac{\delta\Gamma}{\delta\BRSTaBRSTsrcGhost{c}_a}
\right]=0.\hspace{0.5cm}
\label{antiST-id}
\eea

The background Ward identity follows as a consequence of the
local antighost equation and the ST identity, since
\be
0 = {\cal S}_\Gamma ( \bar {\cal G}_a\Gamma  - {\cal D}^{ab}_\mu A^{*\mu}_b -f^{abc}c^*_bc^c ) + {\cal G}_a {\cal S}(\Gamma) = {\cal W}_a \Gamma \, .
\ee
In the above equation ${\cal S}_\Gamma$ is the linearized ST operator
\begin{align}
{\cal S}_\Gamma \equiv  \int\!\diff^4x & \left [
 \frac{\delta\Gamma}{\delta A^{*a}_\mu}\frac{\delta}{\delta A^{\mu}_a}
+\frac{\delta\Gamma}{\delta A^{\mu}_a} \frac{\delta}{\delta A^{*a}_\mu}
+\frac{\delta\Gamma}{\delta c^*_a}\frac{\delta}{\delta c^a}
+\frac{\delta\Gamma}{\delta c^a}\frac{\delta}{\delta c^*_a} \right.&
\nonumber\\
& \left.+\caBRSTsrc{A}{a}_\mu\frac{\delta}{\delta\cBRSTaBRSTsrc{A}{a}_\mu}+\aBRSTsrc{c}_a\frac{\delta}{\delta\BRSTaBRSTsrcGhost{c}_a}+\aBRSTsrc{\bar c}_a\frac{\delta}{\delta\aBRSTsrc{b}_a}+
b^a\frac{\delta}{\delta\bar c^a} 
\right ], &
\end{align}
while the background Ward operator reads
\begin{align}
{\cal W}_a \equiv & -{\cal D}^{ab}_\mu \frac{\delta}{\delta A_{b\mu}}
-\BRSTaBRSTcovD \frac{\delta}{\delta \BRSTaBRSTsrc{A}_{b\mu}} 
+ f^{abc} c_c \frac{\delta}{\delta c_b} 
+ f^{abc} {\bar c}_c \frac{\delta}{\delta {\bar c}_b}
+ f^{abc} b_c \frac{\delta}{\delta b_b} & \nonumber \\
& + f^{abc} \BRSTsrc{c}_c \frac{\delta}{\delta \BRSTsrc{c}_b}
  + f^{abc} \aBRSTsrc{c}_c \frac{\delta}{\delta \aBRSTsrc{c}_b}
  + f^{abc} \aBRSTsrc{\bar c}_c \frac{\delta}{\delta \aBRSTsrc{\bar c}_b}
  + f^{abc} \BRSTsrc{A}_{\mu c} \frac{\delta}{\delta \BRSTsrc{A}_{\mu b}}
  + f^{abc} \aBRSTsrc{A}_{\mu c} \frac{\delta}{\delta \aBRSTsrc{A}_{\mu b}}
& \nonumber \\
& + f^{abc} \BRSTaBRSTsrcGhost{c}_c \frac{\delta}{\delta \BRSTaBRSTsrcGhost{c}_b} 
 + f^{abc} \aBRSTsrc{b}_c \frac{\delta}{\delta \aBRSTsrc{b}_b} \, .
\end{align}

In a similar fashion, the background Ward identity can also be 
obtained by taking the anticommutator between the linearized antiST operator
and the local ghost equation operator ${\cal G}$ and then using
the antiST identity and the local ghost equation.

\subsection{Two-point ghost sector}

The presence of the antighost equation allows to fully constrain the ghost two-point sector in any gauge.
In this sector there are four superficially divergent Green's functions, namely $\Gamma_{c^a\bar c^b}$, $\Gamma_{\bar c^a \Omega^{b}_\mu}$, $\Gamma_{c^aA^{*b}_\mu}$ and, finally, $\Gamma_{\Omega^{a}_\mu A^{*b}_\nu}$ (in the following we prefer to switch back to the familiar notation of $\Omega$ rather than using its antiBRST source name $\aBRSTsrc{A}$). The first two functions are constrained by the ghost equation (we factor out the trivial color structure~$\delta^{ab}$)
\bea
\Gamma_{c\bar c}(q)&=&-iq^\mu\Gamma_{cA^*_\mu}(q),\nonumber \\ 
\Gamma_{\Omega_\mu\bar c}(q)&=&iq_\mu-iq^\nu\Gamma_{\Omega_\mu A^{*}_\nu}(q).
\label{eq12}
\eea

On the other hand, differentiating the antighost equation~(\ref{AGE-xi}) with respect to a gluon anti-field and an antighost, one gets the deformed identities
\bea
\Gamma_{A^*_\mu c}(q)&=&iq_\mu+iq^\nu\Gamma_{A^*_\mu\Omega_\nu }(q)-\xi\Gamma_{A^*_\mu\aBRSTsrc{b} }(q),\nonumber\\
\Gamma_{\bar c c}(q)&=&iq^\mu\Gamma_{\bar c\Omega_\mu}(q)-\xi\Gamma_{\bar c\aBRSTsrc{b}}(q),
\label{eq34}
\eea
and the functions $\Gamma_{\aBRSTsrc{b} A^*_\mu}$ and $\Gamma_{\aBRSTsrc{b} \bar c}$ related through the identity
\be
\Gamma_{\aBRSTsrc{b} \bar c}(q)=-iq^\mu\Gamma_{\aBRSTsrc{b} A^*_\mu}(q).
\label{FP-eta}
\ee

Contracting the first equation in~\noeq{eq34} with $q^\mu$ and next using the first of the identities~\noeq{eq12} as well as~\1eq{FP-eta}, we find the relation 
\be
\Gamma_{c\bar c}(q)=q^2-q^\mu q^\nu\Gamma_{\Omega_\mu A^*_\nu}(q)-\xi\Gamma_{\aBRSTsrc{b}\bar c}(q),
\label{funrel-xi}
\ee
which shows the appearance of the extra function $\Gamma_{\aBRSTsrc{b}\bar c}$ with respect to the Landau gauge, where the ghost sector is entirely determined by  $\Gamma_{\Omega A^*}$ alone.


Then, observing that
\be
\Gamma_{c\bar c}(q)=q^2F^{-1}(q^2),
\ee
where $F$ is the ghost dressing function related to the ghost propagator $D$ through  $D(q^2)=F(q^2)/q^2$, 
and introducing the Lorentz decompositions
\begin{align}
\Gamma_{cA^*_\mu}(q)&=iq_\mu C(q^2);&
\Gamma_{\bar c \Omega_\mu}(q)&=iq_\mu E(q^2);\nonumber \\\Gamma_{\aBRSTsrc{b}\bar c}(q)&=-q^2K(q^2);&
\Gamma_{\Omega_\mu A^*_\nu}(q)&=-g_{\mu\nu}G(q^2)-\frac{q_\mu q_\nu}{q^2}L(q^2),
\label{lorentz-dec}
\end{align}
we finally find the relations
\bea
C(q^2)& =& E(q^2)+\xi K(q^2)\ =\ F^{-1}(q^2),
\nonumber \\
F^{-1}(q^2)&=&1+G(q^2)+L(q^2)+\xi K(q^2).
\label{funrel-formfactors}
\eea
In particular, the last equation above represents the generalization to any $\xi$ of the corresponding well-known identity in the Landau gauge~\cite{Kugo:1995km,Grassi:2004yq,Aguilar:2009pp}; once evaluated at zero momenta, this relation yields the deformation of the Kugo-Ojima confinement criterion~\cite{Kugo:1979gm} in $R_\xi$ gauges.

The function $G$ can be obtained by considering the correlation function corresponding to the time-ordered product of two covariant derivatives, one acting on a ghost and one on an antighost field:
\be
{\cal G}^{ab}_{\mu\nu}(y-x)=\langle T\big[\left({\cal D}^{am}_\nu c^m\right)_x\left({\cal D}^{bn}_\mu \bar c^n\right)_y\big]\rangle=-\frac{\delta W}{\delta\Omega^b_\mu(y)\delta A^{*a}_\nu(x)},
\label{G-connected}
\ee
where $W$ is the generating functional for the connected graphs, see \1eq{legendre} below.

\begin{figure}[!t]
\includegraphics[scale=0.75]{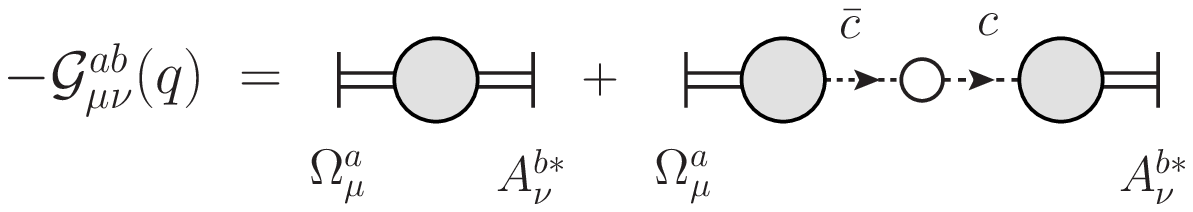} 
\caption{\label{connect}The connected Green's function ${\cal G}_{\mu\nu}$. Grey blobs indicate 1-PI functions, while white ones indicate connected functions (propagators).}
\end{figure}

Now, as shown in \fig{connect}, there are only two possible connected contributions to the Green's function above; using then~\1eq{funrel-formfactors} and passing to a momentum space representation (while factoring out the trivial color structure $\delta^{ab}$), one finds
\bea
{\cal G}_{\mu\nu}(q)&=&-\Gamma_{\Omega_\mu A^*_\nu}(q)-\Gamma_{\Omega_\mu\bar c}(q)D(q)\Gamma_{c A^*_\nu}(q)\nonumber\\
&=&g_{\mu\nu}G(q^2)+\frac{q_\mu q_\nu}{q^2}L(q^2)-\frac{q_\mu q_\nu}{q^2}E(q^2)F(q^2)C(q^2)\nonumber\\
&=&P_{\mu\nu}(q)G(q^2)-\frac{q_\mu q_\nu}{q^2},
\label{G-rel}
\eea
where the transverse projector $P_{\mu\nu}(q)=g_{\mu\nu}-q_\mu q_\nu/q^2$ has been defined.

The important point here is that the relation~\noeq{G-rel} is precisely the same one has in the Landau gauge; therefore  knowledge of the ${\cal G}_{\mu\nu}$ Green's function  translates into a direct determination of the $G$ also in a $R_\xi$ gauge. As the correlator~\noeq{G-connected} is accessible on the lattice, it would be extremely interesting to study its dependence on $\xi$, and in particular determining its behavior as $\xi$ and $q$ go to zero.

\subsection{Two-point gluon sector}

Let us conclude this section by providing a simple proof for the relation~\cite{Kugo:1995km}
\be
\frac{Z_g}{Z_Q}=1+G(0),
\label{kugo}
\ee
where $Z_g$ and $Z_{Q}$ are the charge and (quantum) gauge boson renormalization constants (with a 0 subscript indicating bare quantities)
\be
g=Z_g g_0;\qquad
Q=Z_QQ_0.
\ee
This relation, which is valid for any value of the gauge-fixing parameter,  was first noticed by Kugo~\cite{Kugo:1995km} where however it was proved in a simplified way using classical currents. Below we offer a fully quantum all-order proof.

From the ST identity~\noeq{ST-id} one obtains the relations
\begin{align}
\Gamma_{\cBRSTaBRSTsrc{A}{a}_\mu\cBRSTaBRSTsrc{A}{b}_\nu}(q)&=-\Gamma_{\caBRSTsrc{A}{a}_\mu A^{*c}_\rho}(q)\Gamma_{A^\rho_c \cBRSTaBRSTsrc{A}{b}_\nu}(q),&\nonumber\\
\Gamma_{\cBRSTaBRSTsrc{A}{a}_\mu{A}^{b}_\nu}(q)&=-\Gamma_{\caBRSTsrc{A}{a}_\mu A^{*c}_\rho}(q)\Gamma_{A^\rho_c {A}^{b}_\nu}(q).
\end{align}
Using then the identifications~\noeq{ids}, and reintroducing the background-quantum splitting, one obtains the familiar background-quantum identities~\cite{Grassi:1999tp,Binosi:2002ez}
\begin{align}
\Gamma_{\widehat{A}^{a}_\mu\widehat{A}^{b}_\nu}(q)&=\left[g_{\mu\rho}-\Gamma_{\Omega^{a}_\mu A^{*c}_\rho}(q)\right]\Gamma_{Q^\rho_c \widehat{A}^{b}_\nu}(q),&\nonumber\\
\Gamma_{\widehat{A}^{a}_\mu{Q}^{b}_\nu}(q)&=\left[g_{\mu\rho}-\Gamma_{\Omega^{a}_\mu A^{*c}_\rho}(q)\right]\Gamma_{Q^\rho_c {Q}^{b}_\nu}(q).
\end{align}
Next, we combine the two equations above taking into account the transversality of the two-point gluon function, as well as the Lorentz decomposition~\noeq{lorentz-dec} of the function $\Gamma_{\Omega A^*}$, to get
\be
\Gamma_{\widehat{A}\widehat{A}}(q^2)=[1+G(q^2)]^2\Gamma_{QQ}(q^2),
\ee
where the color ($\delta^{ab}$) and Lorentz ($P_{\mu\nu}$) structures have been factored out.  If we are interested only in the UV part of this identity one can set $q^2=0$, thus obtaining\footnote{Notice that a possible renormalization factor $Z_c$ for $G$ has been entirely reabsorbed in the definition of this quantitiy.}
\be
Z_{\widehat{A}}^{-2}=[1+G(0)]^2Z^{-2}_{Q},
\ee
where we have introduced the background field renormalization constant $\widehat{A}=Z_{\widehat{A}}\widehat{A}_0$. We now take advantage of the residual background gauge invariance which implies the QED-like relation
\be 
Z_{\widehat{A}}^{-1}=Z_g,
\ee
to get finally the desired relation~\noeq{kugo}. 

When originally derived in~\cite{Kugo:1995km} this relation was discussed in the context of the so-called Kugo-Ojima confinement criterion~\cite{Kugo:1979gm}, which predicts that, in the Landau gauge,  a sufficient condition for color confinement is $1+G(0)=0$ (which would in turn imply an IR divergent ghost dressing function). 
It turns out, however, that
lattice data~(see~\cite{Cucchieri:2013nja} for the most recent lattice analysis of the Yang-Mills ghost sector) in conjunction with Schwinger-Dyson techniques~\cite{Aguilar:2009pp}, show that $1+G(0)\neq0$ and thus that there is nothing special about the ratio~\noeq{kugo} (apart obviously the fact that it constitutes a universal, albeit gauge dependent, quantity).

\section{\label{sec:BEA}Background Effective Action}

The requirement of BRST and antiBRST invariance in the presence of scalar and fermionic matter leads to the generalization of the ST and antiST identities of~\2eqs{ST-id}{antiST-id} shown in~\2eqs{ST.full}{aST.full} of Appendix A. The sources of the antiBRST
variations for the gauge, scalar and fermionic matter fields
are to be identified with the corresponding background fields.

In order to further elucidate the physical content
of the ST and antiST identities,
it is convenient to construct an effective action $\widetilde \G$ 
for the background configurations by integrating out completely the quantum
fields.  That is, one is interested in keeping only connected diagrams with external background legs.
 
The functional $\widetilde \G$ , which is therefore one-particle reducible w.r.t. the quantum fields,
can be formally obtained as follows.
The connected generating functional $W$ is obtained by
taking a Legendre transform w.r.t. $\Phi$:
\begin{align}
W = \G + \int \diff^4 x \, J_\Phi \Phi;
\qquad \qquad & J_\Phi = - (-1)^{\epsilon(\Phi)} \frac{\delta \G}{\delta \Phi}; \qquad
\Phi = \frac{\delta W}{\delta J_\Phi}; \nonumber \\
& 
\frac{\delta W}{\delta \zeta} = \frac{\delta \G}{\delta \zeta}; 
\quad \zeta \in \{\BRSTaBRSTsrc{\Phi},\Phi^*, \aBRSTsrc{\Phi}\} , &
\label{legendre}
\end{align}
where we use a collective notation, with $J_\Phi$ denoting the source of the quantum field $\Phi$ and $\epsilon(\Phi)$ the statistics of the field $\Phi$ ($1$ for anticommuting variables, $0$ for commuting ones).

Then one sets
\bea
\widetilde \G[\BRSTaBRSTsrc{\Phi}, \Phi^*, \aBRSTsrc{\Phi}] \equiv \left . W [J_\Phi, \BRSTaBRSTsrc{\Phi}, \Phi^*, \Phi^\#] \right |_{J_{\Phi} = 0}.
\eea
\1eq{ST.full} yields the following identity for the connected functional $W$:
\begin{align}
\!\!\!\!\!
\int\!\diff^4x&\left[ - \frac{\delta W}{\delta A^{*a}_\mu} J_{A^{\mu}_a}
+ \frac{\delta W}{\delta c^*_a}J_{c^a}
+ \frac{\delta W}{\delta J_{b^a}} J_{\bar c^a}
- \frac{\delta W}{\delta{\BRSTsrc{\phi}}^\dagger} J_{\phi} +
\frac{\delta W}{\delta\BRSTsrc{\phi}} J_{\phi^\dagger} +
\frac{\delta W}{\delta \BRSTsrc{\bar\psi}} J_{\psi}
- \frac{\delta W}{\delta \BRSTsrc{\psi}} J_{\bar \psi}\right . &\nonumber \\
&\left.
+ \caBRSTsrc{A}{a}_\mu\frac{\delta W}{\delta\widehat{A}_{a\mu}}
+ \aBRSTsrc{c}_a\frac{\delta W}{\delta\widehat{c_a}}
+\aBRSTsrc{\bar c}_a\frac{\delta W}{\delta\widehat{b_a}}
 -\ \aBRSTsrc{\phi} \frac{\delta W}{\delta \widehat{\phi}}
+ {\aBRSTsrc{\phi}}^\dagger \frac{\delta W}{\delta {\widehat{\phi}}^{\dagger}}
- \aBRSTsrc{\psi}  \frac{\delta W}{\delta \widehat{\psi}}
+ \aBRSTsrc{\bar \psi} \frac{\delta W}{\delta \widehat{\bar \psi}}\right]=0.&
\label{ST.full.W}
\end{align}
By taking a derivative of the above equation 
w.r.t. any of the  antiBRST sources $\aBRSTsrc{\Phi}$ and then
setting all the sources $J_\Phi$ and $\aBRSTsrc{\Phi}$ to zero, one finds that 
\bea
 \frac{\delta \widetilde \G}{\delta \widehat{\Phi}} =
\left . \frac{\delta W}{\delta \widehat{\Phi}} \right |_{J_\Phi =\aBRSTsrc{\Phi}=0} 
=  0 \, .
\label{stat.point}
\eea
This means that the background field configurations $\widehat \Phi$
constitute a stationary point for the background effective action $\widetilde \G$.

Notice that the same result is obtained if one starts from the antiST identity for the connected generating functional $W$, takes one derivative w.r.t. the BRST source $\Phi^*$ and then sets all the sources $J_\Phi$ and $\Phi^*$ to zero.

As a physical example, one can consider the effective field theory of the
Color Glass Condensate (CGC)~\cite{Iancu:2000hn,Ferreiro:2001qy}, which
describes the physics of high gluon densities and gluon saturation in the small $x$-regime ($x$ denoting the longitudinal momentum fraction of the parton in the collision). In this framework, the fulfillment of the ST identity 
for $\widetilde \G$ is crucial for guaranteeing the consistency of the approximations used, as it shows that the background field configuration 
$\widehat A$ is still a stationary point of the background effective
action, even in the presence of radiative corrections induced
by the integration of certain quantum modes of the gluon field.

\subsection{Two- and three-point background gauge functions}

In order to illustrate the combinatorics behind the stationary
condition (\ref{stat.point}), let us consider the case of the two- and
three-point background gauge functions.

\begin{figure}[!t]
\includegraphics[scale=0.65]{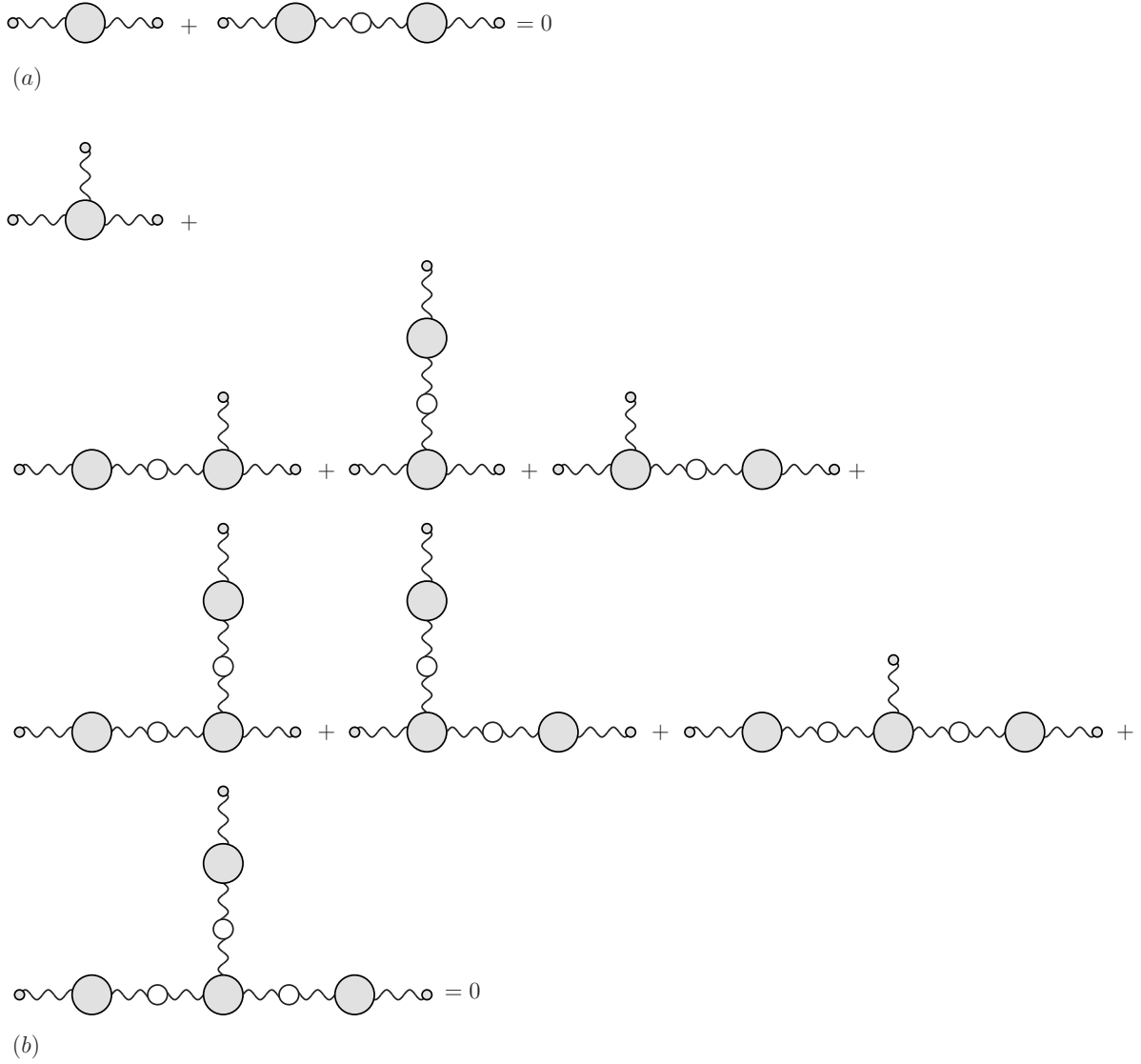} 
\caption{\label{2-3point}The cancellations encoded in the stationary condition~\noeq{stat.point} in the case of the two- and three-point functions of the background gauge field. Small circles attached at the end of lines indicate the background gluons. }
\end{figure} 

The graphs contributing to the two-point function
$\widetilde \G_{\widehat A \widehat A}$ are depicted in~\fig{2-3point}.
From \1eq{legendre} one sees that 
\bea
\int\! \diff^4 z \, \G_{A^{a}_{\mu}A^{c}_ {\rho}} (x,z)W_{J_{A^{c}_{\rho}}  J_{A^{b}_{\nu}}}(z,y) = - \delta_{ab} g^{\mu\nu} \delta^4(x-y) \, .
\label{eq.2pt.gauge}
\eea
If one replaces the 1-PI functions $\G_{\widehat{A} A}$ in the second of the diagrams in \fig{2-3point}$(a)$
by exploiting the background-quantum identity 
\bea
\G_{\widehat A^{a}_{\mu} A^{b}_{\nu}} = - \G_{\Omega^{a}_{\mu} A^{*c}_{\rho}} \G_{A^{b}_{\nu} A^{c}_{\rho}} ,
\label{eq.bkg-quantum.2pt}
\eea
the gauge propagator cancels against one of the 1-PI 2-point
gauge functions by \1eq{eq.2pt.gauge}. The identity 
\be
\widetilde \G_{\widehat A^{a_1}_{\mu_1} \widehat A^{a_2}_{\mu_2}} = 0
\ee
then boils down to the usual background-quantum relation
\bea
\G_{\widehat A^{a_1}_{\mu_1}\widehat A^{a_2}_{\mu_2}}(x_1,x_2) = 
\int\! \diff^4 z_1\int\! \diff^4 z_2 \, 
\G_{\Omega^{a_1}_{\mu_1} A^*_{c_1\rho_1}}(x_1,z_1) 
\G_{\Omega^{a_2}_{\mu_2} A^*_{c_2\rho_2}}(x_2,z_2)
\G_{A_{c_1 \rho_1} A_{c_2 \rho_2}}(z_1,z_2). 
\eea

The identity for the 3-point function
\bea
\widetilde \G_{\widehat A^{a_1}_{\mu_1} \widehat A^{a_2}_{\mu_2} \widehat A^{a_3}_{\mu_3}}(x_1,x_2,x_3) = 0
\label{3point}
\eea
is more involved.
The diagrams contributing to  \1eq{3point} are depicted in \fig{2-3point}$(b)$. They are arranged according to the number of $\widehat A$-insertions
in the 1-PI vertex bubbles. In order to establish the connection
with the 1-PI background-quantum identities, we notice that the legs
involving a gauge propagator and a mixed background-quantum 
amplitude $\G_{\widehat A A}$  can be reduced with the help of 
\1eq{eq.bkg-quantum.2pt} as follows:
\bea
\G_{\widehat A^{a}_{\mu} A^{b}_{\nu}} W_{J_{A^{b}_{\nu}} J_{A^{c}_{\rho}}} = 
- \G_{\Omega^{a}_{\mu} A^{*d}_{\sigma}} \G_{A^{d}_{\sigma} A^{b}_{\nu}} W_{J_{A^{b}_{\nu}} J_{A^{c}_{\rho}}} =
\G_{\Omega^{a}_{\mu} A^{*c}_{\rho}},
\label{ext.leg}
\eea
where \1eq{eq.2pt.gauge} has been used.

After the replacement in \1eq{ext.leg} has been carried out,
one obtains from \1eq{3point}
a representation for $\G_{\widehat A \widehat A \widehat A}$ which only involves 1-PI amplitudes.
It is uniquely determined by the requirement that the ST identity
holds. It can therefore be obtained by applying the method
of canonical transformations presented in~\cite{Binosi:2012st}.
Specifically, by taking the derivative of the ST identity w.r.t. $\Omega$ and setting $\Omega$ to zero afterwards, one obtains quite generally an identity
of the form
\bea
\left . \frac{\delta \G}{\delta \widehat A^{a}_{\mu}} \right |_{\Omega =0} =
- \left . \{ \frac{\delta \G}{\delta \Omega^{a}_{\mu}}, \G \} \right |_{\Omega = 0},
\label{can.t}
\eea
where $\{ \cdot, \cdot \}$ denotes the BV bracket associated with
the ST identity.
\1eq{can.t} states that the dependence on the background
is generated via a canonical transformation with respect
to the BV bracket, induced by the generating functional
$\frac{\delta \G}{\delta \Omega}$.
The latter in general is $\widehat A$-dependent. As a consequence, 
the solution cannot
be written by simple exponentiation of the BV bracket 
w.r.t. $\frac{\delta \G}{\delta \Omega}$, but requires the introduction
of a Lie series of a suitable functional differential operator~\cite{Binosi:2012st}.

For our purposes it is sufficient to consider the reduced bracket
\bea
\{ \frac{\delta \G}{\delta \Omega^{a}_{\mu}(x)}, ~ \cdot ~ \} \equiv 
\int\! \diff^4z \left[
- \G_{\Omega^{a}_{\mu}(x) A^{*b}_{\nu}}(x,z) \frac{\delta}{\delta A^{b}_{\nu}(z)} 
+ \G_{\Omega^{a}_{\mu} A^{b}_{\nu}}(x,z) \frac{\delta}{\delta A^*_{b\nu}(z)}  
\right].
\eea
Then the Lie series generating the background field dependence is 
obtained by exponentiating the operator
\bea
\Delta_{\G_{\Omega^{a}_{\mu}}(x)} \equiv \{ \frac{\delta \G}{\delta \Omega^{a}_{\mu}}, \cdot \} + \frac{\delta}{\delta \widehat A^{a}_{\mu}},
\eea
that is, one has
\bea
\G  =  E_{\G_\Omega} (\G_0) + \cdots,
\label{lie}
\eea
where the dots denote  amplitudes involving at least one 
external leg different than $A,\widehat A$ and the mapping $E_{\G_\Omega}$ is defined according to
\bea
E_{\G_\Omega} (\G_0)
 \equiv \sum_{n \geq 0} \frac{1}{n!}\int_1 \dots \int_n \widehat A_1 \dots \widehat A_n 
[\Delta_{\G_{\Omega_1}} \dots \Delta_{\G_{\Omega_n}} \G_0]_{\widehat A = 0} \, .
\label{lie.series}
\eea
In the above equation $\G_0$ denotes the 1-PI vertex functional where fields and sources
have been set to zero, with the only exception of $A,A^*$; finally,  the shorthand notations
$\int_i = \int\! \diff^4 z_i$,  $\widehat A_i = \widehat A^{a_i}_{\mu_i}(z_i)$ 
and $\G_{\Omega_i} = \G_{\Omega^{a_i}_{\mu_i}}(z_i)$ have been used.

We stress that \1eq{lie}  reproduces the correct dependence of the
1-PI vertex functional on $\widehat A$ in the gauge sector
only. The background dependence of amplitudes 
involving scalar and fermionic matter fields as well as other external sources can be recovered by making use of the full
canonical transformation generated by the functional $\frac{\delta \G}{\delta \Omega}$.

The first two terms of \1eq{lie} yield (notice that 
in the  following equations $\widehat A$ is set equal to zero, while further
differentiations w.r.t. $A$ are possible):
\be
\G_{\widehat A^{a_1}_{\mu_1}}(x_1)  = - \int\! \diff^4 z_1\,
\G_{\Omega^{a_1}_{\mu_1} A^{*b_1}_{\beta_1}}(x_1,z_1) \G_{A^{b_1}_{\beta_1}}(z_1) ,
\label{1.point.can}
\ee
and
\begin{align}
%
\G_{\widehat A^{a_1}_{\mu_1}\widehat A^{a_2}_{\mu_2}}(x_1,x_2) 
 =
\frac{1}{2} \int\! \diff^4 z_1\int\!\diff^4 z_2 \, &\left [ 
\G_{\Omega^{a_2}_{\mu_2} A^{*b_2}_{\beta_2}}(x_2,z_2)
\G_{\Omega^{a_1}_{\mu_1} A^{*b_1}_{\beta_1} A^{b_2}_{\beta_2}}(x_1,z_1,z_2) \G_{A^{b_1}_{\beta_1}}(z_1) \right.
\nonumber \\
&+
\G_{\Omega^{a_2}_{\mu_2} A^{*b_2}_{\beta_2}}(x_2,z_2)
\G_{\Omega^{a_1}_{\mu_1} A^{*b_1}_{\beta_1}}(x_1,z_1) \G_{A^{b_2}_{\beta_2} A^{b_1}_{\beta_1}}(z_2,z_1)
 \nonumber \\
& \left.   - \G_{\Omega^{a_1}_{\mu_1} A^{*b_1}_{\beta_1}\widehat A^{a_2}_{\mu_2}}(x_1,z_1,x_2) \G_{A^{b_1}_{\beta_1}}(z_1) \right]  + (a_1\mu_1 \leftrightarrow a_2\mu_2) \, ,
\label{2.point.can}
\end{align}
while the third term gives for the three point function
$\G_{\widehat A\widehat A\widehat A}$
(we suppress the space-time arguments):
\begin{align}
\G_{\widehat A^{a_1}_{\mu_1} \widehat A^{a_2}_{\mu_2} \widehat A^{a_3}_{\mu_3} }
 = 
\frac{1}{3!} 
\int\! \diff^4 z_1\int\! \diff^4 z_2\int\!  \diff^4 z_3  & \left[  
- \G_{\Omega^{a_3}_{\mu_3} A^{*b_3}_{\beta_3}} 
  \G_{\Omega^{a_2}_{\mu_2} A^{*b_2}_{\beta_2}}
  \G_{\Omega^{a_1}_{\mu_1} A^{*b_1}_{\beta_1} A^{b_2}_{\beta_2}} 
  \G_{A^{b_3}_{\beta_3} A^{b_1}_{\beta_1}} \right. \nonumber \\
& 
- \G_{\Omega^{a_3}_{\mu_3} A^{*b_3}_{\beta_3}} 
  \G_{\Omega^{a_2}_{\mu_2} A^{*b_2}_{\beta_2} A^{b_3}_{\beta_3}}
  \G_{\Omega^{a_1}_{\mu_1} A^{*b_1}_{\beta_1} } 
  \G_{A^{b_2}_{\beta_2} A^{b_1}_{\beta_1}}  \nonumber \\
&
- \G_{\Omega^{a_3}_{\mu_3} A^{*b_3}_{\beta_3}} 
  \G_{\Omega^{a_2}_{\mu_2} A^{*b_2}_{\beta_2} }
  \G_{\Omega^{a_1}_{\mu_1} A^{*b_1}_{\beta_1} A^{b_3}_{\beta_3}} 
  \G_{A^{b_2}_{\beta_2} A^{b_1}_{\beta_1}}  \nonumber \\
&
- \G_{\Omega^{a_3}_{\mu_3} A^{*b_3}_{\beta_3}} 
  \G_{\Omega^{a_2}_{\mu_2} A^{*b_2}_{\beta_2} }
  \G_{\Omega^{a_1}_{\mu_1} A^{*b_1}_{\beta_1}} 
  \G_{ A^{b_3}_{\beta_3} A^{b_2}_{\beta_2} A^{b_1}_{\beta_1}}  \nonumber \\
& 
+ \G_{\Omega^{a_3}_{\mu_3} A^{*b_3}_{\beta_3}}  
  \G_{\Omega^{a_1}_{\mu_1} A^{*b_1}_{\beta_1} \widehat A^{a_2}_{\mu_2}} 
  \G_{A^{a_3}_{\mu_3} A^{b_1}_{\beta_1}} \nonumber \\
& 
+ \G_{\Omega^{a_2}_{\mu_2} A^{*b_2}_{\beta_2} \widehat A^{a_3}_{\mu_3}}
  \G_{\Omega^{a_1}_{\mu_1} A^{*b_1}_{\beta_1}} \G_{A^{b_2}_{\beta_2} A^{b_1}_{\beta_1}}
\nonumber \\
&\left.
+ \G_{\Omega^{a_2}_{\mu_2} A^{*b_2}_{\beta_2}} \G_{\Omega^{a_1}_{\mu_1} A^{*b_1}_{\beta_1} \widehat A^{a_3}_{\mu_3}} \G_{A^{b_2}_{\beta_2} A^{b_1}_{\beta_1}} \right]  + \mathrm{symm.}
\label{3.point.bkg}
\end{align}
where complete symmetrization of the r.h.s.
of the above equation w.r.t. the $a_i,\ \mu_i$ indices is understood.
Compatibility of the diagrammatic identity in~\fig{2-3point}$(b)$ with \1eq{3.point.bkg} follows by taking the appropriate derivatives w.r.t. the
quantum fields $A$ of \2eqs{1.point.can}{2.point.can} in order to eliminate
recursively the background insertions in the amplitudes
of the second and third lines of~\fig{2-3point}$(b)$. 

Notice in particular the  presence
of the amplitudes $\G_{\Omega A^* \widehat A}$. The latter
arise due to the dependence of the generating functional
 $\G_{\Omega}$ on the background $\widehat A$. 
The amplitudes $\G_{\Omega A^* \widehat A}$ can be fully fixed
neither by the ST nor the antiST identities.
Their $\widehat A$-dependence
is the cause of the failure of the simple exponentiation in order
to derive the solution to \1eq{can.t}, which in turn
can be overcome by using the appropriate Lie series in
\1eq{lie.series}.

\section{\label{sec:concl}Conclusions}

In this paper we have shown that the ($R_\xi$) BFM naturally emerges once the requirement of BRST and antiBRST invariance of the action is fulfilled: indeed, background fields are unequivocally identified as the sources associated to the operator $s\,\bs$.
Correspondingly the existence of the antiBRST symmetry implies background gauge invariance (and, consequently, a  background Ward identity) as well as a (new) local antighost equation, which, when used in conjunction with the  local ghost equation enforced by the BRST symmetry, completely determines the algebraic structure of the ghost sector of the theory for any gauge-fixing parameter. In addition, the background fields have been shown to be an extremum of the background effective action obtained by integrating out the quantum fields.

In hindsight,  the correspondence
\be
\mathrm{BRST}+\mathrm{antiBRST}\equiv\mathrm{BFM},
\label{literal}
\ee 
might not appear all that unexpected, as in the BFM background sector the ghost trilinear vertex is proportional to the sum of the ghost and antighost momentum while a quartic vertex involving two background fields and a ghost and an antighost (proportional to the metric tensor) is also generated. Thus the ghost and antighost are treated in a symmetric fashion, exactly as required by the BRST and antiBRST invariance.
It should be stressed however that~\1eq{literal} works only in the $R_\xi$ gauges (which is anyway the only practically relevant case); choosing, e.g., a non-covariant background gauge breaks irremediably  the antiBRST symmetry of the theory.

In Chapter~15 of the second volume of his ``Quantum theory of fields''~\cite{Weinberg:1996kr} S. Weinberg noticed: ``{\it The discovery of invariance under an `antiBRST' symmetry showed that, despite appearances, there is a similarity between the roles of [the ghost field] $\omega^A$ and [the antighost field] $\omega^{*A}$ which remains somewhat mysterious.}''  

We hope that this paper helps to shed some light on the mystery.

\acknowledgments
One of us (A.~Q.) thanks the European Centre for Theoretical Studies in Nuclear Physics and Related Areas (ECT*) where part of this work has been carried out.

\appendix

\section{Functional identities in the presence of  scalar and fermionic 
matter fields\label{app:A}}

We give here the relevant functional identities in the presence of scalar
and fermionic matter fields.
The ST identity takes the form
\begin{align}
\int\!\diff^4x&\left[\frac{\delta\Gamma}{\delta A^{*a}_\mu}\frac{\delta\Gamma}{\delta A^{\mu}_a}+\frac{\delta\Gamma}{\delta c^*_a}\frac{\delta\Gamma}{\delta c^a}+\caBRSTsrc{A}{a}_\mu\frac{\delta\Gamma}{\delta\cBRSTaBRSTsrc{A}{a}_\mu}+\aBRSTsrc{c}_a\frac{\delta\Gamma}{\delta\BRSTaBRSTsrcGhost{c}_a}
+\aBRSTsrc{\bar c}_a\frac{\delta\Gamma}{\delta\aBRSTsrc{b}_a}
+b^a\frac{\delta\Gamma}{\delta\bar c^a} \right .&\nonumber \\
&+ \frac{\delta\Gamma}{\delta{\BRSTsrc{\phi}}^\dagger} \frac{\delta \Gamma}{\delta \phi} -
\frac{\delta \Gamma}{\delta\BRSTsrc{\phi}} \frac{\delta \Gamma}{\delta \phi^\dagger} +
\frac{\delta \Gamma}{\delta \BRSTsrc{\bar\psi}} \frac{\delta \Gamma}{\delta \psi}
- \frac{\delta \Gamma}{\delta \BRSTsrc{\psi}} \frac{\delta \Gamma}{\delta \bar \psi}&\nonumber \\
&\left. -\ \aBRSTsrc{\phi} \frac{\delta \Gamma}{\delta \BRSTaBRSTsrc{\phi}}
+ {\aBRSTsrc{\phi}}^\dagger \frac{\delta \Gamma}{\delta {\widetilde{\phi}^{\#\dagger}}} 
- \aBRSTsrc{\psi}  \frac{\delta \Gamma}{\delta \BRSTaBRSTsrc{\psi}}
+ \aBRSTsrc{\bar \psi} \frac{\delta \Gamma}{\delta \BRSTaBRSTsrc{\bar \psi}}\right]=0.&
\label{ST.full}
\end{align}

The antiST identity is
\begin{align}
\int\!\diff^4x&\left[\frac{\delta\Gamma}{\delta \caBRSTsrc{A}{a}_\mu}\frac{\delta\Gamma}{\delta A^{\mu}_a}+\frac{\delta\Gamma}{\delta \aBRSTsrc{c}_a}\frac{\delta\Gamma}{\delta c^a}+\frac{\delta\Gamma}{\delta \aBRSTsrc{\bar c\hspace{0.05cm}}_a}\frac{\delta\Gamma}{\delta \bar c^a}+\frac{\delta\Gamma}{\delta \aBRSTsrc{b}_a}\frac{\delta\Gamma}{\delta b^a}-A^{*a}_\mu\frac{\delta\Gamma}{\delta\cBRSTaBRSTsrc{A}{a}_\mu}-c^{*}_a\frac{\delta\Gamma}{\delta\BRSTaBRSTsrcGhost{c}_a}\right.& \nonumber \\
& 
+ \frac{\delta\Gamma}{\delta{\aBRSTsrc{\phi}}^\dagger} \frac{\delta \Gamma}{\delta \phi} -
\frac{\delta \Gamma}{\delta\aBRSTsrc{\phi}} \frac{\delta \Gamma}{\delta \phi^\dagger} +
\frac{\delta \Gamma}{\delta \aBRSTsrc{\bar\psi}} \frac{\delta \Gamma}{\delta \psi}
- 
\frac{\delta \Gamma}{\delta \aBRSTsrc{\psi}} \frac{\delta \Gamma}{\delta \bar \psi} &
\nonumber \\
& \left . 
 +\ \BRSTsrc{\phi} \frac{\delta \Gamma}{\delta \BRSTaBRSTsrc{\phi}}
- {\BRSTsrc{\phi}}^\dagger \frac{\delta \Gamma}{\delta {\widetilde{\phi}^{\#\dagger}}} 
+ \BRSTsrc{\psi}  \frac{\delta \Gamma}{\delta \BRSTaBRSTsrc{\psi}}
- \BRSTsrc{\bar \psi} \frac{\delta \Gamma}{\delta \BRSTaBRSTsrc{\bar \psi}}
\right]=0.&
\label{aST.full}
\end{align}
Proceeding in the same way as for the derivation of
\2eqs{AGE-xi}{GE-xi}, we obtain the
local antighost equation
\bea
\frac{\delta\Gamma}{\delta c^a} &=&-f^{abc}\frac{\delta\Gamma}{\delta b^b}\bar c^c-\xi\frac{\delta\Gamma}{\delta \aBRSTsrc{b}_a}+\BRSTaBRSTcovD\frac{\delta\Gamma}{\delta \caBRSTsrc{A}{b}_\mu}+{\cal D}^{ab}_\mu A^{*\mu}_b+f^{abc}c^*_bc^c+f^{abc}\BRSTaBRSTsrcGhost{c}_b\frac{\delta\Gamma}{\delta \aBRSTsrc{c}_c}+f^{abc}\aBRSTsrc{b}_b\frac{\delta\Gamma}{\delta \aBRSTsrc{\bar c\hspace{0.05cm}}_c}
\nonumber \\
& + & i \frac{\delta \Gamma}{\delta \aBRSTsrc{\phi}} t^a 
\BRSTaBRSTsrc{\phi} 
+ i {\widetilde{\phi}}^{\#\dagger} t^a \frac{\delta \Gamma}{\delta 
{\aBRSTsrc{\phi}}^\dagger}
+ i \frac{\delta \Gamma}{\delta \aBRSTsrc{\psi}} t^a 
\BRSTaBRSTsrc{\psi}
+ i \BRSTaBRSTsrc{\bar \psi} t^a \frac{\delta \Gamma}{\delta \aBRSTsrc{\bar\psi}} \nonumber \\
&-&  i \phi^\dagger t^a \BRSTsrc{\phi} - i {\BRSTsrc{\phi}}^\dagger t^a \phi
+ i \bar \psi t^a \BRSTsrc{\psi} + i \BRSTsrc{\bar \psi} t^a \psi,
\label{AGE-xi.full}
\eea
and the ghost equation
\bea
\frac{\delta\Gamma}{\delta \bar c^a}& = &{\cal D}^{ab}_\mu \caBRSTsrc{A}{\mu}_b-\BRSTaBRSTcovD\frac{\delta\Gamma}{\delta A^{*b}_\mu}-f^{abc}\BRSTaBRSTsrcGhost{c}_b\frac{\delta\Gamma}{\delta c^*_c}+f^{abc} \aBRSTsrc{c}_b c^c+f^{abc}\aBRSTsrc{\bar c\hspace{0.05cm}}_b\bar c^c-f^{abc}\aBRSTsrc{b}_bb^c
\nonumber \\
& -& 
 i \frac{\delta \Gamma}{\delta \BRSTsrc{\phi}} t^a 
\BRSTaBRSTsrc{\phi} 
- i {\widetilde{\phi}}^{\#\dagger} t^a \frac{\delta \Gamma}{\delta 
{\BRSTsrc{\phi}}^\dagger}
- i \frac{\delta \Gamma}{\delta \BRSTsrc{\psi}} t^a 
\BRSTaBRSTsrc{\psi}
- i \BRSTaBRSTsrc{\bar \psi} t^a \frac{\delta \Gamma}{\delta \BRSTsrc{\bar\psi}} \nonumber \\
&-&  i \phi^\dagger t^a \aBRSTsrc{\phi} - i {\aBRSTsrc{\phi}}^\dagger t^a \phi
+ i \bar \psi t^a \aBRSTsrc{\psi} + i \aBRSTsrc{\bar \psi} t^a \psi.
\label{GE-xi.full}
\eea

Finally the $b$ equation becomes
\bea
\frac{\delta\Gamma}{\delta b^a} & = &\BRSTaBRSTcovD(A^\mu_b-\cBRSTaBRSTsrc{A}{\mu}_b)-\xi b^a-f^{abc}\aBRSTsrc{b}_b\bar c^c-\aBRSTsrc{c}_a-f^{abc}\BRSTaBRSTsrcGhost{c}_bc^c,
\nonumber \\&+&i\cBRSTaBRSTsrc{\phi}{\dagger}t^a\phi-i\BRSTaBRSTsrc{\phi}t^a\phi^\dagger+i\BRSTaBRSTsrc{\bar \psi}t^a\psi-i\BRSTaBRSTsrc{\psi}t^a\bar \psi,
\label{bE.full}
\eea
while the background Ward identity yields
\begin{align}
{\cal W}_a \Gamma = & -{\cal D}^{ab}_\mu \frac{\delta \Gamma}{\delta A_{b\mu}}
-\BRSTaBRSTcovD \frac{\delta \Gamma}{\delta \BRSTaBRSTsrc{A}_{b\mu}} 
+ f^{abc} c_c \frac{\delta \Gamma}{\delta c_b} 
+ f^{abc} {\bar c}_c \frac{\delta \Gamma}{\delta {\bar c}_b}
+ f^{abc} b_c \frac{\delta \Gamma}{\delta b_b} & \nonumber \\
&
 + f^{abc} \BRSTsrc{c}_c \frac{\delta \Gamma}{\delta \BRSTsrc{c}_b}
  + f^{abc} \aBRSTsrc{c}_c \frac{\delta \Gamma}{\delta \aBRSTsrc{c}_b}
  + f^{abc} \aBRSTsrc{\bar c}_c \frac{\delta \Gamma}{\delta \aBRSTsrc{\bar c}_b}
  + f^{abc} \BRSTsrc{A}_{\mu c} \frac{\delta \Gamma}{\delta \BRSTsrc{A}_{\mu b}}
  + f^{abc} \aBRSTsrc{A}_{\mu c} \frac{\delta \Gamma}{\delta \aBRSTsrc{A}_{\mu b}}
&
 \nonumber \\
 & + f^{abc} \BRSTaBRSTsrcGhost{c}_c \frac{\delta \Gamma}{\delta \BRSTaBRSTsrcGhost{c}_b} 
 + f^{abc} \aBRSTsrc{b}_c \frac{\delta \Gamma}{\delta \aBRSTsrc{b}_b} 
 + i t^a \phi \frac{\delta \Gamma}{\delta \phi} 
 - i \phi^\dagger t^a \frac{\delta \Gamma}{\delta \phi^\dagger}
 + i t^a \psi \frac{\delta \Gamma}{\delta \psi}
 - i \bar \psi t^a \frac{\delta \Gamma}{\delta \bar \psi} 
& = 0 \, .
\end{align}


\end{document}